# Weak correlation between fluctuations in protein diffusion inside bacteria

Yuichi Itto[1,2] and Christian Beck[3]

[1] Science Division, Center for General Education, Aichi Institute of Technology,
    Aichi 470-0392, Japan

[2] ICP, Universität Stuttgart, 70569 Stuttgart, Germany

[3] School of Mathematical Sciences, Queen Mary University of London,
    London E1 4NS, UK

**Abstract**. A weak correlation between the diffusion-exponent fluctuations and the temperature fluctuations is discussed based on recent experimental observations for protein diffusion inside bacteria. Its existence is shown to be essential for describing the statistical properties of the fluctuations. It is also quantified how largely the fluctuations are modulated by the weak correlation.

# 1. Introduction

A recent experiment in Ref. [1] (see also Ref. [2]) has elucidated a highly heterogeneous diffusion phenomenon for histonelike nucleoid-structuring proteins (i.e., DNA-binding proteins) in living *Escherichia coli* bacteria. Such nucleoid-associated proteins interact with DNA as well as themselves, distributing uniformly over the bacterium. The experimental technique of single-particle tracking has made intriguing results available at the level of individual trajectories of the proteins: the mean square displacement of a given protein, $\overline{\Delta x^2}$, scales as

$$\overline{\Delta x^2} \sim D_\alpha t^\alpha \qquad (1)$$

for elapsed time, $t$, where $D_\alpha$ is the diffusion constant and $\alpha$ is the (anomalous) diffusion exponent. (Here and hereafter, the symbols we use are slightly different from those in Ref. [1].) Normal diffusion shows $\alpha = 1$, whereas the case with $\alpha \neq 1 (>0)$ corresponds to anomalous diffusion [3-5].

An exotic observation [1] is that not only the diffusion constants but also the diffusion exponents fluctuate in a wide range: the distribution of the diffusion constants decays as the following power law

$$\phi(D_\alpha) \sim D_\alpha^{-\gamma-1} \qquad (2)$$

with the exponent $\gamma \cong 0.97$, whereas the diffusion exponent obeys a non-trivial distribution in the range $0 \leq \alpha \leq 2$. It is noted that the distribution in Eq. (2) has been obtained for dimensionless numerical values of $D_\alpha$ since the dimension of $D_\alpha$ changes depending on the values of $\alpha$. For small elapsed time, only normal diffusion has been observed, for which the diffusion constant, $D$, also obeys asymptotically a power law, $\phi(D) \sim D^{-\mu-1}$ with the exponent $\mu \cong 1.9$. The power-law nature of the fluctuations is an unexpected result, since it is quite different from the exponential law found for bacteria, for example, in Ref. [6] (see also Refs. [7,8] for an entropic approach to this law and Ref. [9] for a formal analogy of such a fluctuating diffusivity to thermodynamics).

The distributions of fluctuations are known to play a central role for describing non-Gaussian distributions of displacements, a point which will be mentioned later in the present context. In fact, there are relevant discussions suitable for biological application: in Refs. [10,11], the diffusion-constant fluctuations are treated, whereas the diffusion-exponent fluctuations are considered in Refs. [12,13], (which were originally motivated by a work in Ref. [14]). These discussions are based on the concept of so-called superstatistics [15] (see Refs. [16-27] for recent developments, for example). In addition, the fluctuations similar to the present ones have experimentally been observed in other biological systems as reported in Refs. [28-31].

A remarkable feature [1] is as follows. The bacteria have been classified into three groups based on their cell age (or, equivalently cell length), for which the mean square displacement in an ensemble average, i.e., an average of square displacement over all of the individual trajectories has been analyzed, giving rise to both an average diffusion constant and an average diffusion exponent. The experimental results for such groups have then shown that in terms of the cell age, the average diffusion constant increases significantly, whereas the average diffusion exponent is approximately constant, i.e., it increases *only slightly*.

In this article, we discuss a weak correlation between the diffusion-exponent fluctuations and the temperature fluctuations developed in a recent work in Ref. [32]. It is shown that the existence of such a weak correlation is essential for describing the statistical properties of the fluctuations. We also quantify how largely the fluctuations are modulated by the weak correlation.

## 2. Weak correlation

Let us consider that the proteins exhibit a stochastic motion over the bacterium, which is regarded as a complex medium for the protein diffusion and is divided into many small local regions or "blocks". There, both the diffusion constant and the diffusion exponent in Eq. (1) slowly fluctuate locally, the time scale of which is much larger than that of the dynamics of the proteins. In non-equilibrium statistical physics, the diffusion constant is proportional to temperature through the Einstein relation [33], in general. Therefore, we focus ourselves on the joint fluctuations of the diffusion exponent, $\alpha$, and the inverse temperature, $\beta$, given without loss of generality by

$$g(\alpha, \beta) = g(\alpha | \beta) f(\beta), \qquad (3)$$

where $g(\alpha | \beta)$ is the conditional fluctuation distribution describing the probability of $\alpha$, given a value of $\beta$, and $f(\beta)$ is the marginal fluctuation distribution describing the probability of $\beta$.

For $f(\beta)$, we suppose the following $\chi^2$ distribution [34]:

$$f(\beta) = \frac{1}{\Gamma(\mu)} \left( \frac{\mu}{\beta_0} \right)^{\mu} \beta^{\mu-1} \exp\left( -\frac{\mu \beta}{\beta_0} \right) \qquad (4)$$

in the whole range of $\beta \in (0, \infty)$, where $\beta_0$ is the average value of $\beta$ and $\Gamma(\mu)$ is the Euler gamma function. Under the assumption of the relation $D \propto 1/\beta$, Eq. (4) turns out to offer an inverse gamma distribution in terms of $D$, decaying asymptotically as $\phi(D)$, which fits with the experimental data quite well, as can be seen in Fig. 3 in Ref. [32]. In addition, the distribution of the dimensionless numerical values of $D_\alpha$ has also been found to take the form of an inverse gamma distribution, exhibiting the power-law nature of Eq. (2), see Fig. 1 in Ref. [32].

Regarding $g(\alpha | \beta)$, let us recall the experimental fact that the average diffusion exponent increases only slightly with respect to the cell age. This fact suggests the existence of a *weak* correlation between $\alpha$ and $\beta$ at the statistical level. In other words, $\alpha$ and $\beta$ are not fully statistically independent. Accordingly, we describe the conditional fluctuation distribution based on the weak correlation. To do so, let us express it as

$$g(\alpha | \beta) = e^{h(\alpha | \beta)} \qquad (5)$$

with a suitable function, $h(\alpha|\beta)$, which is approximately constant in terms of $\beta$ in its whole range: its first derivative with respect to $\beta$ is small so that we expand it around at $\beta = \beta_0$ up to the first order of $\beta - \beta_0$, i.e., $h(\alpha|\beta) \cong h_0(\alpha) + h_1(\alpha)(\beta - \beta_0)$, where $h_0(\alpha) \equiv h(\alpha|\beta_0)$ and $h_1(\alpha) \equiv h'(\alpha|\beta_0)$, and the prime stands for differentiation with respect to $\beta$. Therefore, we have the following conditional fluctuation distribution:

$$g(\alpha|\beta) \sim g(\alpha|\beta_0)\exp[(\beta - \beta_0)h_1(\alpha)] \qquad (6)$$

with $h_1(\alpha)$ being small, guaranteeing the weakness of correlation.

From Eqs. (4) and (6), we can immediately obtain the marginal fluctuation distribution $g(\alpha) = \int d\beta\, g(\alpha, \beta)$, leading to

$$g(\alpha) \sim g(\alpha|\beta_0)\frac{\exp(-\beta_0 h_1(\alpha))}{[1 - (\beta_0/\mu)h_1(\alpha)]^\mu}, \qquad (7)$$

where the quantity $1 - (\beta_0/\mu)h_1(\alpha)$ has been assumed to be positive and this is indeed the case for Eq. (12) below, see Ref. [32] for this point.

Equation (7) is formal without determining the conditional distribution. Below, we propose a concrete form for it. Like the Einstein relation, for $D_\alpha$ in Eq. (1), we assume the following relation

$$D_\alpha = D_{\alpha,\beta} \sim \frac{c}{s^\alpha \beta}, \qquad (8)$$

where $s$ is the characteristic time being required for displacement of the protein and $c$ is a positive constant. This relation seems to be experimentally supported as discussed in Ref. [32] (see also Ref. [35] for a relevant discussion). Motivated by the above-mentioned distribution of $D_\alpha$, it is supposed [32] that given a value of $\beta$, the normalized probability distribution of the dimensionless numerical values of $D_\alpha$ also follows the inverse gamma distribution given by

$$\phi(D_\alpha) \propto \tilde{A}^\gamma D_\alpha^{-\gamma-1} \exp\left(-\frac{\gamma \tilde{A}}{D_\alpha}\right) \tag{9}$$

with $\tilde{A} = \tilde{A}(\beta)$ being a dimensionless positive quantity depending on $\beta$. From Eqs. (8) and (9), therefore, the conditional fluctuation distribution is given by

$$g(\alpha \mid \beta) = \left|\frac{\partial D_{\alpha,\beta}}{\partial \alpha}\right| \phi(D_\alpha), \tag{10}$$

where $\alpha$ is supposed to be distributed in the interval $0 \leq \alpha \leq 2$. Since this distribution should have a weak dependence on $\beta$, $\tilde{A}(\beta)$ is considered to take the following form: $\tilde{A}(\beta) = a(\beta)/\beta$, where $a(\beta)$ is a positive quantity depending weakly on $\beta$ and accordingly is expanded as $a(\beta) \cong a_0 + a_1(\beta - \beta_0)$ with $a_0 \equiv a(\beta_0)$ and $a_1 \equiv a'(\beta_0)$. To realize the weakness, $a_1$ should be small. Thus, the conditional fluctuation distribution behaves as

$$g(\alpha \mid \beta) \propto s^{\gamma\alpha} \exp\left(-\frac{\gamma a(\beta)}{c} s^\alpha\right), \tag{11}$$

from which $h_1(\alpha)$ is found to be given by

$$h_1(\alpha) = \frac{\gamma a_1}{c}(\langle s^\alpha \rangle_\alpha - s^\alpha) \tag{12}$$

with $\langle Q \rangle_\alpha \equiv \int_0^2 d\alpha\, g(\alpha|\beta_0) Q,$ see Ref. [32] for details. It is noted here that the distribution in Eq. (11) is peaked at $\alpha = \alpha_*(\beta) = [\ln(c/a(\beta))]/\ln s,$ and the condition $\alpha'_* < 0$ is imposed since the center of this distribution should tend to approach the origin $\alpha = 0$ as $\beta$ increases, which turns out to require $a_1$ to be negative. As can be seen in Fig. 2 in Ref. [32], $g(\alpha)$ in Eq. (7) with Eqs. (11) and (12) shows a good agreement with the experimental data, highlighting how essential the existence of the weak correlation is.

## 3. Quantification of the modulation of fluctuations by weak correlation

Equation (7) shows that the conditional fluctuation distribution is modulated by the weak correlation. In this section, we quantitatively discuss such a modulation by introducing the following quantity:

$$G(\alpha|\beta) = \ln \frac{g(\alpha)}{g(\alpha|\beta)}. \tag{13}$$

This gives a measure about how largely the fluctuations are modulated by the weak correlation: $G > 0$ ($G < 0$), if the modulation is enhanced (suppressed). The quantity is calculated to be

$$G(\alpha|\beta) \sim -\mu \ln\left(1 - \frac{\beta_0}{\mu} h_1(\alpha)\right) - \beta h_1(\alpha), \tag{14}$$

where Eqs. (6) and (7) have been used.

In Fig. 1, we present the plot of $G(\alpha|\beta)$ in Eq. (14) with Eq. (12) in the case of $\beta = \beta_0$. There, it is observed that the quantity takes large values near $\alpha = 0$, in particular, meaning that the modulation is dominant for small values of $\alpha$. This seems to explain why a local minimum in the marginal distribution $g(\alpha)$ appears (see Fig. 2 in Ref. [32]), showing an interesting aspect of the role of the weak correlation.

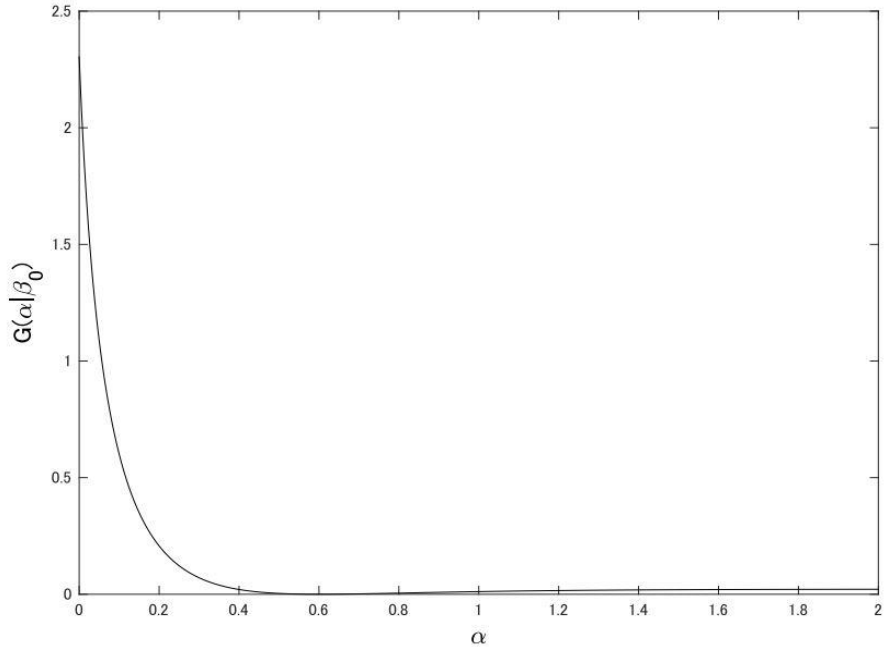

**Figure 1.** The quantity $G(\alpha\,|\,\beta)$ in Eq. (14) with Eq. (12) at $\beta = \beta_0$. The same numerical values as employed in Ref. [32] are used. All quantities are dimensionless.

## 4. Concluding remarks

We have discussed a weak correlation between the diffusion-exponent fluctuations and the temperature fluctuations based on a highly heterogeneous diffusion phenomenon observed experimentally for histonelike nucleoid-structuring proteins in living *Escherichia coli* bacteria. We have shown that the existence of the weak correlation is crucial for describing the statistical properties of the fluctuations. We have also quantified how largely the fluctuations are modulated by the weak correlation.

As mentioned in the Introduction, fluctuations play a central role for describing the non-Gaussian displacement distributions. For the experimental results in Ref. [1], the displacements of the proteins have been found to obey a $q$-Gaussian distribution [36] (also called a Pearson-type VII distribution [37]), which exhibits a power-law behavior for large displacements. In Ref. [32], such a distribution has been derived by incorporating, in a superstatistical way, the fluctuations into a process of fractional Brownian motion [38,39], for which a fractional operator [40] constitutes a key, in accordance with the experimental data. In Ref. [41], fluctuating diffusivity of proteins is discussed in terms of protein conformational fluctuation, which is very relevant in our context.

**Acknowledgements**

The present article is based on a talk given by Y. I. at the 10th International Conference on Mathematical Modeling in Physical Sciences (September 6-9, 2021, Virtual, on-line Conference). He would like to thank the organizers of the conference for providing him with the opportunity to give a talk. The authors thank Ralf Metzler for drawing their attention to Ref. [41]. The work of Y. I. has been supported by a Grant-in-Aid for Scientific Research from the Japan Society for the Promotion of Science (No. 21K03394).